\documentclass{article}

\usepackage{setspace,graphics}
\usepackage[dvips]{epsfig} 
\usepackage{amssymb,epsfig,array,cite,setspace,fontenc,times,mathptmx,graphicx}
\usepackage{slashed}
\usepackage{amsmath}

\newcounter{multieqs}

\newcommand{\be}{\begin{equation}}
\newcommand{\ee}{\end{equation}}
\newcommand{\eq}[1]{(\ref{#1})}

\def\nn{\nonumber}
\def\bea{\begin{eqnarray}}
\def\eea{\end{eqnarray}}
\def\obar{\overline}

\def\beqa{\begin{eqnarray}} 
\def\eeqa{\end{eqnarray}} 
\def\beq{\begin{equation}} 
\def\eeq{\end{equation}}

\def\Tr{{\rm Tr}}

\def\a{\alpha}

\def\d{\delta}

\def\cA{{\cal A}}  \def\cC{{\cal C}}
 \def\cE{{\cal E}} 
 \def\cH{{\cal H}} 
  
\def\cM{{\cal M}}

\def\R{{\mathbb R}}

\def\one{\mbox{1 \kern-.59em {\rm l}}}

\def\bit{\begin{itemize}}
\def\eit{\end{itemize}}

\def\({\left(}
\def\){\right)}

\def\d{\delta}

\def\uno{\mbox{1 \kern-.59em {\rm l}}}

\newcommand{\tr}{\mbox{tr}}

\def\bcomment#1{}

 \def\ii{{\rm i}}

\newcommand{\ad}{\mathrm{d}}
\newcommand{\lek}{\left[}
\newcommand{\rek}{\right]}
\newcommand{\lrk}{\left(}
\newcommand{\rrk}{\right)}

\newcommand{\omu}{^{\mu}}

\newcommand{\onu}{^{\nu}}

\newcommand{\omunu}{^{\mu\nu}}

\newcommand{\bsp}{\begin{split}}
\newcommand{\esp}{\end{split}}

\newcommand{\thetainv}{\theta^{-1}}

\renewcommand{\title}[1]{\vspace{10mm}\noindent{\Large{\bf #1}}\vspace{8mm}}
\newcommand{\authors}[1]{\noindent{\large #1}\vspace{5mm}}
\newcommand{\address}[1]{{\itshape #1\vspace{2mm}}}

\begin{document}

\begin{titlepage}

\begin{center}
  
\title{Fermions Coupled to Emergent Noncommutative Gravity} \\
\vspace{2cm}
%\today

\authors{Daniela {\sc Klammer}${}^{1}$ and Harold {\sc Steinacker}${}^{2}$}

 \address{ Fakult\"at f\"ur Physik, Universit\"at Wien\\
 Boltzmanngasse 5, A-1090 Wien, Austria 
 }

%\footnotetext[1]{harald.grosse@univie.ac.at}
\footnotetext[1]{daniela.klammer@univie.ac.at, Talk given at the Workshop on Black Holes in General Relativity and String Theory, 24 - 30 August, 2008, Veli Lo\^ sinj, Croatia.}
\footnotetext[2]{harold.steinacker@univie.ac.at}

\vskip 2cm

\textbf{Abstract}

\vskip 3mm 

\begin{minipage}{14cm}%%                  SET-UP
We study the coupling of fermions to Yang-Mills matrix models in the framework of emergent noncommutative gravity. It is shown that the matrix model action provides an appropriate coupling for fermions to gravity, albeit with a non-standard spin-connection. Integrating out the fermions in a nontrivial geometrical background induces indeed the Einstein-Hilbert action for on-shell geometries plus a dilaton-like term. This result explains UV/IR mixing as a gravity effect. It also illuminates why UV/IR mixing remains even in supersymmetric models, except in the $N=4$ case.
\end{minipage}

\end{center}

\end{titlepage}

%\setcounter{page}0
%\thispagestyle{empty}
%\newpage

%%%%%%%% table of content %%%%%%%
\begin{spacing}{.3}
{
\noindent\rule\textwidth{.1pt}            % THEN MAKE TOC...
   \tableofcontents
\vspace{.6cm}
\noindent\rule\textwidth{.1pt}
}
\end{spacing}

%%%%%%%ordinary document (end) ####################################

\section{Introduction}

\paragraph{Noncommmutativity \& gravity.} Heisenberg's uncertainty principle together with Einsteins' general theory of relativity lead to the conclusion that the classical concept of spacetime loses its meaning in the small. When measuring a spacetime coordinate with great accuracy $a$, there is an uncertainty in momentum of the order $1/a$. That is to say measuring small distances requires high energies, which will curve locally the region of spacetime you want to measure. When the gravitational field becomes so strong as to prevent any signal from escaping that region the operational meaning of this localization gets lost. The process of measuring a spacetime coordinate to infinite accuracy is thus as a matter of principle not possible. 

It has been shown in a fundamental paper by Doplicher, Fredenhagen and Roberts~\cite{Doplicher:1994tu} that the above argument leads to uncertainty relations for the spacetime coordinates which can be derived from \emph{noncommuting spacetime coordinates}, such as 
\begin{align}
\lek x\omu, x\onu \rek = \ii \theta\omunu.
\end{align}
One can then study so called called ``noncommutative (NC) quantum field theories'' on spaces with such noncommuting coordinates; for basic reviews see e.g.~\cite{Szabo:2001kg,Douglas:2001ba}. 
In NC field theories quantum fluctuations of spacetime coordinates occur naturally. Thus it is believed that these theories could play an important r\^ole on the way towards a quantum theory of gravity. Recently, a specific realization of this idea was published under the name of ``emergent noncommutative gravity,'' see ~\cite{Steinacker:2007dq} and ~\cite{Grosse:2008xr,Klammer:2008df,Steinacker:2008ri,Steinacker:2008ya}. There, matrix models of noncommutative gauge theory describe dynamical noncommutative spaces. The main lesson learned is that gravity is already contained in noncommutative gauge theories. There is no need to add new concepts. Here we discuss specific results of this approach: We study the successful coupling of fermions to the framework of emergent noncommutative gravity.

\section{Matrix models and effective geometry}
\label{sec:metric}

Consider the matrix model action 
\be
S_{YM} = - Tr [Y^a,Y^b] [Y^{a'},Y^{b'}] g_{a a'} g_{b b'}
\label{YM-action-1}
\ee
for
\be
g_{a a'} = \delta_{a a'} \quad \mbox{or}\quad g_{a a'} = \eta_{a a'} 
\label{background-metric}
\ee
in the Euclidean  resp.  Minkowski case. $g_{a a'}$ should not be interpreted as a fixed, physical background metric, but rather as a prescription to fix the signature.  
Here the ``covariant coordinates'' $Y^a$ for $a=1,2,3,$ are hermitian matrices 
or operators acting on some Hilbert space $\cH$.
We will denote the commutator of two matrices as 
\be
[Y^a,Y^b] = i \theta^{ab}
\ee
so that $\theta^{ab} \in L(\cH)$ is an antihermitian matrix, which is \emph{not} assumed to be constant here. 
We study configurations $Y^a$ (not necessarily solutions of 
the equation of motion) which can be interpreted as quantizations 
of a Poisson manifold $(\cM,\theta^{ab}(y))$ with general
Poisson structure $\theta^{ab}(y)$. This defines the geometrical 
background under consideration, and 
essentially any (local) Poisson manifold is a possible background $Y^a$.
In particular, we assume that 
$\theta^{ab}$ is small and well approximated by the classical
Poisson structure $\theta^{ab}(y)$ in a semi-classical expansion. 
More formally, this means that there is an isomorphism of vector spaces
\be
\cC(\cM) \to \cA \subset L(\cH)\, 
\ee
where $\cC(\cM)$ denotes the space of functions on $\cM$,
and $\cA$ is the algebra generated by $Y^a$,
interpreted as quantized algebra of functions. 
In particular, $Y^a$ corresponds to a classical coordinate function\footnote{The coordinates $y^a$ are preferred ones since in their frame $g_{a a'}$ equals $\delta_{a a'}$ resp. $\eta_{a a'}$. In other frames $g_{a a'}$ will not be constant.}
$y^a$ on $\cM$. This can be used to define a star product on $\cC(\cM)$.
Moreover, $Y^a$ defines a derivation on $\cA$ via
\begin{align}
\lek Y^a, f\rek \sim \ii \theta^{ab}(y) \partial_b f(y).
\end{align}

In this paper, we  restrict ourselves to the ``irreducible'' case,
i.e. we assume that $\cA$ is in some sense dense in $L(\cH)$. 
Then any matrix (``function'') in $L(\cH)$  
can be expressed as a function of $Y^a$ resp. $y^a$.
From the gauge theory point of view in Section \ref{change-of-variables}, it
means that we restrict ourselves to the $U(1)$ case.
This is most interesting for us since the UV/IR mixing (see Sect. \ref{sec:UV-IR-mixing}) happens in the
trace-$U(1)$ sector. 
For the general case see \cite{Steinacker:2007dq,Steinacker:2008ya}.

\paragraph{Scalars.}

To begin with, we consider the case of scalar fields 
i.e. hermitian matrices $\Phi$ coupled to 
the matrix model \eq{YM-action-1}. There it is seen most easily how the effective metric appears. 
The only possibility 
to write down kinetic terms for matter fields is through commutators 
$[Y^a,\Phi] \sim i\theta^{ab}(y) \frac{\partial}{\partial y^b} \Phi$\footnote{Throughout this paper, $\sim$ 
indicates the leading contribution in a semi-classical 
expansion in powers of $\theta^{ab}$.},
and one is lead to the action
\bea
S[\Phi] = (2\pi)^2\, \Tr\, g_{aa'} [Y^a,\Phi][Y^{a'},\Phi]  
\sim \int d^4 y\, \rho(y)\, G^{ab}(y)\,\frac{\partial}{\partial y^a}\Phi(y) 
\frac{\partial}{\partial y^b} \Phi(y) .
\label{scalar-action-0}
\eea
Here 
\bea\label{effective-metric}
G^{ab}(y)=\theta^{ac}(y)\theta^{bd}(y)g_{cd}
\eea
is the effective metric for the scalar field $\Phi$. The Poisson manifold naturally acquires a metric structure 
$(\cM, \theta^{ab}(y), G^{ab}(y))$ determined by the Poisson structure.  The metric is thus no fundamental building block of the theory. We also used $\Tr \sim \int \ad^4 y \rho(y)$, where 
\be
\rho(y) =  (\det\theta^{ab}(y))^{-1/2} = |G_{ab}(y)|^{1/4}  
\equiv e^{-\sigma}
\quad (\,\equiv \Lambda_{NC}^4(y)\,)
\ee
is the symplectic measure on $(\cM,\theta^{ab}(y))$
which can be naturally interpreted as non-commutative 
scale $\Lambda_{NC}^4$.
After appropriate rescaling of $G^{ab}(y)$, this can be 
rewritten in covariant form 
\be
S[\Phi] = \int d^4 y\, \tilde G^{ab}(y) \partial_{y^a}\Phi \partial_{y^b}\Phi 
\label{scalar-action-geom}
\ee
with the effective metric 
\bea
\tilde G^{ab} = |G_{ab}|^{1/4}\, G^{ab} = \rho(y)\, G^{ab}, \qquad 
|\tilde G^{ab}| = 1\, 
\label{metric-unimod}
\eea
being unimodular in the preferred $y^a$ coordinates.

\paragraph{Fermions.}

The most obvious action
for a spinor which can be written down in the matrix 
model framework\footnote{In particular, 
fermions should also be in the adjoint, otherwise they
cannot acquire a kinetic term. This does not rule out its applicability
in particle physics, see e.g. \cite{Steinacker:2007ay}.} is
\be
S = (2\pi)^2\, \Tr  \obar\Psi  \gamma_a [Y^a,\Psi]
 \,\, \sim\,\, \int d^4 y\, \rho(y)\, \obar\Psi i \gamma_a \theta^{ab}(y)
\partial_b \Psi
 \label{fermionic-action-geom}
\ee
ignoring possible nonabelian gauge
fields here to simplify the notation. This is written for the case of
Minkowski signature, the Euclidean version involves the obvious
replacement $\bar\Psi \to \Psi^\dagger$.
This defines the (matrix) Dirac operator
\be
\not \!\! D \,\Psi = \gamma_a [Y^a,\Psi]\, 
\,\, \sim\,\, i \gamma_a \theta^{ab}(y)\partial_b \Psi.
\label{Diracop-matrix}
\ee 
We can compare this with the 
standard covariant derivative for spinors 
\be
\not \!\! D_{\rm comm} \, \Psi 
= i\gamma^a\, e_a^\mu \(\partial_\mu + \Sigma_{ab}\, \omega^{ab}_\mu\)\Psi 
\label{covar-spinor}
\ee
where  
\be
\omega^{ab}_\mu = i\frac 12\, e^{a\nu} \(\nabla_\mu e^b_{\nu}\)
\label{spinconnection-2}
\ee
is the spin connection, and $\Sigma_{ab}=\ii/4 [\gamma_a,\gamma_b]$ is the representation of the Lorentz algebra.
Comparing \eq{Diracop-matrix} with \eq{covar-spinor},
we observe again that in the 
geometry defined by \eq{effective-metric},
\be
e^\mu_b(y) := \theta^{\mu c}(y) g_{cb} 
\label{vielbein}
\ee
plays the r\^ole of a preferred vielbein. However this must
be used with great care, because the distinction between 
the coordinate index $\mu$ and the Lorentz index $a$ is lost
in the special ``gauge'' inherent in \eq{vielbein}.

One notices that 
the spin connection does not appear in
\eq{fermionic-action-geom}, 
which seems very strange 
at first. In spite of this strange feature, the action \eq{fermionic-action-geom}
is a good action for a fermion propagating in the geometry
defined by $\tilde G_{ab}$. To see this, 
recall that the spin connection determines how the spinors are rotated 
under parallel transport along a trajectory.
However, the spin-connection $\omega^{ab}_\mu$ can 
always be eliminated 
(via parallel-transport resp. a suitable gauge choice)
along an open trajectory. Then the conventional
kinetic term \eq{covar-spinor} 
boils down to \eq{fermionic-action-geom}.
Therefore in the point-particle limit,
the trajectory of a fermion with action
\eq{fermionic-action-geom} 
will follow properly the geodesics 
of the metric\footnote{For massless particles, the geodesics of 
$\tilde G_{ab}$ coincide with those of $G_{ab}$.
Masses should be generated spontaneously, which is not considered
here.} $\tilde G_{ab}$, 
albeit with a different rotation
of the spin. Furthermore,  the
induced gravitational action obtained by integrating out the 
fermion in \eq{fermionic-action-geom}
indeed induces the expected Einstein-Hilbert term
$\int d^4 y\,  R[\tilde G]\, \Lambda^2$ at least
for ``on-shell geometries'', albeit with an 
unusual numerical coefficient and an extra term 
depending on $\sigma$.
All this shows that \eq{fermionic-action-geom} defines a
reasonable action for fermions in the background defined by 
$\tilde G_{ab}$.

\paragraph{Equations of motion.} So far we considered arbitrary background configurations $Y^a$ as long
as they admit a geometric interpretation.
The equations of motion derived from the action \eq{YM-action-1} 
\be\label{eom}
[Y^a,[Y^{a'},Y^b]]\, g_{a a'} = 0 \quad \stackrel{\mathrm{semi-cl.\; limit}}{\rightarrow} \quad
\theta^{ma}\partial_m \theta^{nb}g_{ab}=0  
\ee
select on-shell geometries among all possible backgrounds, such as 
the Moyal-Weyl quantum plane \eq{Moyal-Weyl}. 
However since we are interested in the quantization
here, we will need general off-shell configurations below.

\section{Quantization and induced gravity}\label{sec: quantization}

Next we study the quantization of our matrix model coupled to fermions.
In principle, the quantization is defined in terms of a (``path'') integral
over all matrices $Y^a$ and $\Psi$. 
In 4 dimensions, we can only perform perturbative computations
for the ``gauge sector'' $Y^a$, while the fermions can be integrated
out formally in terms of a determinant. Let us focus here on the
effective action at one loop\footnote{For the sake of rigor we work in Euclidean case now.} obtained by integrating out the fermionic fields,
\bea
e^{-\Gamma_{\Psi}} = \int \ad\Psi\ad\bar{\Psi} e^{-S[\Psi]} 
\quad \mathrm{with}\quad 
\Gamma_{\Psi} = -\frac 12 \Tr \log \not \!\! D^2 \, .
\label{trlog}
\eea
for a non-interacting fermionic field with action Eq. \ref{fermionic-action-geom}.

\paragraph{Square of the Dirac operator and induced action.}
The square of the  Dirac operator takes the following form
\bea
\not \!\! D^2 \Psi = \gamma_a\gamma_b [Y^a,[Y^b,\Psi]] 
\sim -\gamma_a\gamma_b \theta^{ac} \partial_c (\theta^{bd} \partial_d \Psi) 
=  - G^{cd} \partial_c\partial_d  \Psi
  -  a^d\partial_d \Psi,
\label{D-square-1}
\eea
with
\bea
a^d &=& \gamma_a\gamma_b \theta^{ma}\partial_m \theta^{db} =
-2i\,\Sigma_{ab}\theta^{ac}\partial_c\theta^{bd} + g_{ab}\theta^{ac}\partial_c\theta^{bd}.
\label{a-def}
\eea
$\not \!\! D^2$ defines the quadratic form
\bea\label{S-square}
S_{\rm square} = (2\pi)^2\, \Tr \Psi^\dagger\not \!\! D^2 \Psi 
\sim \int d^4y\, \rho(y) \Psi^\dagger\not \!\! D^2 \Psi 
=  \int d^4y\, |G_{ab}|^{1/4} \Psi^\dagger\not \!\! D^2 \Psi,
\eea
which is very similar to the scalar action.
In terms of the unimodular metric $\tilde G_{ab}$ Eq. \eq{metric-unimod}, $S_{\rm square}$ 
can be written in standard covariant form
\bea
S_{\rm square} = \int d^4y\, \sqrt{|\tilde G|}\,\,  \obar\Psi\,
\widetilde{\not \!\!  D^2}\Psi \quad \mathrm{with}\quad
\widetilde{\not \!\! D^2} \Psi = -\left(
\tilde G^{cd} \partial_c\partial_d  \Psi
  + e^{-\sigma}\,a^d \partial_d \Psi  \right).
\label{D-squared}
\eea
We now compute the effective action using
\bea
\frac 12 \Tr (\log \widetilde{\not \!\! D^2} - \log \widetilde{\not \!\! D_0^2}) 
&=& -\frac 12 \Tr \int_{0}^\infty d\a \frac 1\a 
\(e^{-\a  \,\widetilde{\not \! D^2}} - e^{-\a  \,\widetilde{\not \! D_0^2}}\)e^{-\frac{1}{2\alpha\widetilde{\Lambda}^2}} ,
\label{heatkernel-expand}
\eea
where $\widetilde{\Lambda}^2$ denotes the cutoff for 
$\widetilde{\not \!\! D^2}$ regularizing the divergence 
for small $\alpha$. Now we can apply the heat kernel expansion
\be
\Tr\, e^{-\alpha \;\widetilde{\not \! D^2}} 
=
\sum_{n\geq 0}\lrk\frac{\alpha}{2}\rrk^{\frac{n-4}{2}} \int_{\cM}d^4 y \;a_n\left(y,\widetilde{\not \!\! D^2}\right)
\ee
where the Seeley-de Witt coefficients $a_n (y,\widetilde{\not \!\! D^2})$ are given by~\cite{Gilkey:1995mj}
\bea
a_0(y) &=& \frac{1}{16\pi^2}\,\tr\,\one,\nn\\
a_2(y) &=& \frac{1}{16\pi^2}\tr\left(\frac{R[\widetilde{G}]}{6}\;\one + \cE \right), \nn\\
\cE &=& - \widetilde{G}^{mn}\left(
\partial_m \Omega_n + \Omega_m\Omega_n -\widetilde{\Gamma}^k_{mn}\Omega_k
\right),  \label{E-def}\\
\Omega_m &=& \frac{1}{2}\widetilde{G}_{mn}\left(e^{-\sigma}a^n +
  \widetilde{\Gamma}^n \right), \label{Omega-def}
\eea
where $\tr$ denotes the trace over the spinorial matrices.
The effective action is therefore
\bea \label{induced gravity action}
\Gamma_{\Psi} &=& \frac 1{16\pi^2}\, \int d^4 y \,\( 2\, \tr(\one)\,\widetilde{\Lambda}^4 
 + \tr \left(\frac{R[\tilde G]}{6}\; \one +\cE\right)\,
 \widetilde{\Lambda}^2 + O(\log \tilde \Lambda) \)\, ,
 \label{S-oneloop-fermions}
\eea
where $\tr (\one) =4$ assuming Dirac fermions.
Everything is expressed in terms of the unimodular metric
$\tilde{G}_{ab}$, 
which can be written in terms of $G_{ab}$ using
\bea
R[\widetilde{G}]&=& \rho(y)\left( R[G]+3\Delta_G \sigma -\frac 32 G^{ab}\partial_a \sigma \partial_b \sigma \right), \nn\\
\Delta_G \sigma &=& G^{ab}\partial_a \partial_b \sigma - \Gamma^c \partial_c \sigma, \nn\\
\Gamma^a &=& G^{bc}\Gamma^a_{bc}, \nn\\
e^{-\sigma(y)}&=&\rho(y)=\lrk\det G_{ab}\rrk^{1/4}, \nn\\
\widetilde{\Gamma}^a&=& \widetilde{G}^{cd}\widetilde{\Gamma}^a_{cd}
 \, =\,e^{-\sigma}\Gamma^a - e^{-\sigma}\lrk\partial_b\sigma\rrk G^{ba}.
\eea
Note the relative minus sign of the various terms in the 
effective action $\Gamma_{\Psi}$ compared with the 
induced action due to a scalar field \cite{Grosse:2008xr}, 
\be
\Gamma_{\Phi}=\frac{1}{16\,\pi^2}\int d^4 y
\left(-2\widetilde{\Lambda}^4 
-\frac{1}{6} R[\widetilde{G}] \widetilde{\Lambda}^2 
+O(\log \widetilde{\Lambda})
\right).
\label{Gamma-phi}
\ee
hence
\be
\Gamma_\Psi + 4\, \Gamma_\Phi 
\,\, =\,\,  \frac{1}{16\,\pi^2}\int d^4 y \,\tr\, \cE \, \widetilde{\Lambda}^2 \,.
\label{induced-susy-E}
\ee
This expresses the cancellation of the induced actions due to 
fermions and bosons, apart from the $\cE$ term.
For the standard coupling of Dirac fermions to gravity
on commutative spaces, one has~\cite{Vassilevich:2003xt}
\be\label{cE-commutative}
\tr\,\cE_{\rm comm} = - R  
\ee
which originates from an additional constant term $-\frac 14 R$ in 
$\not \!\! D^2_{\rm comm}$ (Lichnerowicz's formula).
In our case, $\cE$ turns out to be somewhat modified
due to the missing spin connection, nevertheless it 
contains the appropriate curvature scalar plus an additional 
term, see Eq. \eq{trE-result}.
\paragraph{Ricci scalar in terms of $\theta^{mn}$.}
The curvature is given as usual by
\be
{R_{abc}}^d = \partial_b \Gamma^{d}_{ac} - \partial_a\Gamma^{d}_{bc}
 + \Gamma^{e}_{ac}\Gamma^{d}_{eb} - \Gamma^{e}_{bc}\Gamma^{d}_{ea}\, .
\ee
The Ricci scalar is then
\bea
R= G^{ac}\,R_{abc}^b= G^{ac}\left(\partial_b \Gamma_{ac}^b - \partial_a\Gamma^b_{bc}
+\Gamma^e_{ac}\Gamma^b_{eb}-\Gamma^e_{bc}\Gamma^b_{ea}\right).
\eea
By plugging in the explicit formula for metric tensor,
\be
G^{mn}(y)=\theta^{ma}(y)\theta^{nb}(y) g_{ab}
\ee
one can express the Ricci scalar $R$ in terms of $\theta$. 
By making use of the Jacobi identity,
\bea\label{eq: Jacobi id}
\partial_a \theta^{-1}_{bc}+\partial_c \theta^{-1}_{ab}+\partial_b \theta^{-1}_{ca}=0\\
\partial_a \theta^{pq}=-\lrk \partial_c \theta^{-1}_{am}\rrk\lrk \theta^{mp}\theta^{cq}-\theta^{mq}\theta^{cp}\rrk,
\eea
and by exploiting relations coming from partial integration
several terms appearing in the computation of $\tr\,\cE$ and 
$R[\widetilde{G}]$ are equivalent\footnote{By means of these 
relations one can also check that the action (\ref{S-square}) 
is indeed hermitian.}.
After a rather lengthy computation, which can be found in~\cite{Klammer:2008df}, one yields the following compact form for the Ricci scalar in terms of the unimodular metric $\widetilde{G}_{ab}$ 
\bea
\int d^4 y\, R[\widetilde{G}]\widetilde{\Lambda}^2&=&e^{-\sigma} \Big\{
\frac{1}{2}G^{mk}\lrk\partial_k\thetainv_{na}\rrk G^{nl}\lrk\partial_l\thetainv_{mb}\rrk g^{ab}
-\frac{1}{2}G^{mn}G^{pq}\lrk\partial_p\thetainv_{ma}\rrk\lrk\partial_q\thetainv_{nb}\rrk g^{ab} \nn\\
&\quad&
-\frac{1}{2}\lrk \partial_p \theta^{pa}\rrk G^{qk}\lrk\partial_k \thetainv_{qa}\rrk
+\frac{1}{2} G^{mn}\lrk\partial_m \sigma\rrk\lrk\partial_n \sigma\rrk 
\Big\}\widetilde{\Lambda}^2.
\label{R-action-explicit}
\eea
%%%%%%%%%%%%%%%%%%%%%%%%%%%% Tr \cE %%%%%%%%%%%%%%%%%%%%%%%%%%%%%%%%%%%%%%%%%%%
\paragraph{Evaluation of $\tr\,\cE$.} \label{Tr E}
We also need to evaluate 
\bea
\tr \cE&=&-\tr\widetilde{G}^{ab}\left(\partial_a \Omega_b + \Omega_a\Omega_b - \widetilde{\Gamma}^r_{ab}\Omega_r\right), 
\eea
where
\bea \label{omega-ausruck}
\Omega_m&=&\frac{1}{2}\widetilde{G}_{mn}\left(\widetilde{a}^n+\tilde{\Gamma}^n\right)\nn\\
&=&
\frac{1}{2}\left(
G_{mn}\gamma_a\gamma_b\theta^{pa}\left(\partial_p \theta^{nb}\right)
-G_{mn}\lrk\partial_p G^{pn}\rrk +\partial_m \sigma
\right)
\eea
and $\tilde a^n = e^{-\sigma}\, a^n$.
For the computation of $\tr\,\cE$ we use again the Jacobi
identity (\ref{eq: Jacobi id}) and relations from partial integration and we find:
\bea \label{TrE}
\tr\,\cE &=&
e^{-\sigma} \Big\{
G^{kl}G^{mn}\lrk\partial_k\thetainv_{ma}\rrk\lrk\partial_l\thetainv_{nb}\rrk g^{ab} 
-G^{mk}\lrk\partial_k\thetainv_{na}\rrk G^{nl}\lrk\partial_l\thetainv_{mb}\rrk g^{ab}
\Big\}.
\eea
Comparing with \eq{R-action-explicit}
for $\widetilde{\Lambda}^2$ regarded as constant cutoff 
of $\Delta_{\widetilde{G}}$, we can write this as
\bea
\tr\,\cE &=& -2\,R[\tilde G]
-\lrk \partial_p \theta^{pa}\rrk G^{qk}\lrk\partial_k \thetainv_{qa}\rrk
+ G^{mn}\lrk\partial_m \sigma\rrk\lrk\partial_n \sigma\rrk \nn\\
&\stackrel{\mathrm{eom}}{=}& -2R[\tilde G]+G^{mn}\partial_m\sigma\partial_n\sigma,
\label{trE-result}
\eea
assuming on-shell geometries \eq{eom} in the last line.
This formula applies for Dirac fermions, and 
with an additional factor $\frac 12$ for Weyl fermions.
It is remarkable that $\tr\,\cE$ is essentially given 
by the appropriate 
curvature scalar $R[\tilde G]$, and up to a contribution from the 
dilaton-like scaling factor $\rho = e^{-\sigma}$.  This is a very
reasonable modification of the standard result 
\eq{cE-commutative}, as desired and tells us that Einstein-Hilbert action also emerges for fermions at one-loop.

%%%%%%%%%%%%%%%%%%%%%%%%%%%%%%%%%%%%%%%%%%%%%%%%%%%

\section{Relation with gauge theory 
on $\mathbb{R}_{\theta}^4$}
\label{change-of-variables}

One motivation to study NC field theories comes from the fact that NC spacetime coordinates in the small tend to cure the UV divergencies.  However, the supposedly removed divergencies reappear in the infrared limit $p \rightarrow 0$. This effect is the notorious \emph{UV/IR mixing} which spoils renormalizability~\cite{Minwalla:1999px}. 
In the framework of emergent noncommutative gravity the UV/IR mixing problem of noncommutative gauge theories is understood in terms of an induced gravity action. In order to show this we 
want to interpret the fermionic action \eq{fermionic-action-geom} as 
action for a Dirac fermion on the Moyal-Weyl quantum
plane $\R^4_\theta$
coupled to a $U(1)$ gauge field in the adjoint. 
This point of view
is obtained by writing the general covariant coordinate
resp. matrix $Y^a$ as
\be
Y^a = X^{a} + \cA^a\, .
\label{cov-coord-1}
\ee
Here $X^{a}$ are generators of the Moyal-Weyl quantum plane, which 
satisfy
\be
[X^a,X^b] = i \bar\theta^{ab}\, ,
\label{Moyal-Weyl}
\ee
where $\bar \theta^{ab}$ is a \emph{constant} antisymmetric tensor. 
These are particular
 solutions of the equations of motion \eq{eom}.
The effective geometry 
for the Moyal-Weyl plane is flat, given by
\bea
\bar g^{ab} &=& \bar\theta^{ac}\,\bar\theta^{bd} g_{cd}\, \nn\\
\tilde g^{ab} &=& \bar \rho\, \bar g^{ab} ,
\qquad \det \tilde g^{ab} =1  \nn\\
\bar\rho &=&  (\det\bar\theta^{ab})^{-1/2}
= |\bar g_{ab}|^{1/4} \equiv \Lambda_{NC}^4 \, .
\label{effective-metric-bar}
\eea
Consider now the
change of variables
\be
\cA^a(x) = -\bar\theta^{ab} A_b(x)
\label{A-naive}
\ee
where $A_a$ are hermitian matrices interpreted  as smooth functions 
on $\R^4_{\bar \theta}$. 
 Thus we can write
\bea
\lek Y^a, f \rek=\lek X^a + \cA^a, f\rek = i \bar{\theta}^{ab}\lrk \frac{\partial }{\partial x^b}f + i\lek A_b, f\rek \rrk 
\equiv i\bar{\theta}^{ab} D_b f,
\eea
giving for the quadratic form \eq{S-square}
\bea
S_{square}&=& (2\pi)^2\,  \Tr\; \Psi^\dagger \gamma_a\gamma_b\lek Y^a,
\lek Y^b, \Psi\rek \rek \nn\\
&=&-\int d^4x \,\bar{\rho} \,\Psi^\dagger\;\gamma_a \gamma_b 
\bar{\theta}^{am}\bar{\theta}^{bn}D_m D_n \Psi \nn\\
&=&\int d^4 x\, \Psi^\dagger \,\widetilde{\not \!\! D^2}_A \Psi\, . 
\eea
This is an exact expression on $\R^4_\theta$,
where
\be
\widetilde{\not \!\! D^2}_A 
= - \bar{\rho} \,\gamma_a\gamma_b\bar{\theta}^{am}\bar{\theta}^{bn}D_m D_n  \, 
= -  \,\tilde\gamma^m\tilde\gamma^n\, D_m D_n  \, ,
\ee
and
\be
\tilde \gamma^a = (\det \bar g_{ab})^{\frac 18}\,\gamma_b\, \bar\theta^{ba},   \qquad
\{\tilde \gamma^a,\tilde\gamma^b\} =  2 \,\tilde g^{ab}\, .
\label{tilde-gamma}
\ee
We now want to rewrite the geometrical results of Section \ref{sec: quantization}
in terms of gauge theory on $\R^4_\theta$ in $x$-coordinates. To do this, 
note that most formulas of Section \ref{sec: quantization}
are not generally covariant, but only valid in the preferred
$y$-coordinates defined by the matrix models where
$g_{ab} = \d_{ab}$ resp. $g_{ab} = \eta_{ab}$.
Eq. \eq{cov-coord-1} defines the 
leading-order relation between $y$ and $x$ coordinates, 
\be
y^a = x^a - \bar{\theta}^{ab} \bar{A}_b + O(\theta^2)\, .
\ee
See ~\cite{Klammer:2008df} for details of this change of variables. Let us moreover denote $\bar{\partial}_a = \partial/\partial x^a$. 
The Poisson tensor can be written 
in terms of the $\mathfrak{u}(1)$ field strength as 
\be
i \theta^{ab}(y)=\lek Y^a, Y^b\rek 
= i\bar{\theta}^{ab} - i\bar{\theta}^{ac}\bar{\theta}^{bd}\bar{F}_{cd},
\ee
where $\bar{F}_{cd}$ is a rank two tensor in $x$ coordinates
 on $\mathbb{R}_{\theta}^4$.
We also need the effective metric \eq{effective-metric} in $x$-coordinates, 
 \bea
G^{ab}=\lrk \bar{\theta}^{ac}-\bar{\theta}^{ai}\bar{\theta}^{cj}\bar{F}_{ij}\rrk
 \lrk \bar{\theta}^{bd}-\bar{\theta}^{be}\bar{\theta}^{df}\bar{F}_{ef}\rrk g_{cd}. 
\eea
We find for the one-loop induced action \bea
\Gamma_{\Psi}&=&\int d^4 y \left(
a_0 \widetilde{\Lambda}^4 + a_2 \widetilde{\Lambda}^2 + O\left(\log \widetilde{\Lambda}\right)\right)
\nn\\
&=&-4 \Gamma_{\Phi}-\frac{1}{16\,\pi^2}\int d^4y \frac{\rho(y)}{2}\bar{g}^{ac}\bar{g}^{bd}\bar{F}_{ab}\bar{\partial}^2 \bar{F}_{cd} \,
\widetilde{\Lambda}^2.
\eea
Finally, there is a nontrivial relation between the
cutoff $\widetilde{\Lambda}$ of the geometrical action
and the cutoff $\Lambda$ of the $\mathfrak{u}(1)$ gauge theory,
which follows from the identity
\bea
S_{\rm square}=\Tr\; \Psi^\dagger\gamma_a\gamma_b \lek Y^a,\lek Y^b,\Psi\rek \rek 
= \int d^4y\,\Psi^\dagger\widetilde{\not \!\! D^2}_{\widetilde{G}}\Psi   %\nn\\
= \int d^4 y \frac{\rho(y)}{\bar{\rho}}\Psi^\dagger\widetilde{\not \!\! D^2}_A \Psi.
\eea
For the Lapacians this means
\be
\widetilde{\not \!\! D^2}_{\widetilde{G}} =\frac{\rho(y)}{\bar{\rho}}\widetilde{\not \!\! D^2}_A.
\ee
Since we implement the cutoffs using Schwinger parameterization they are related as follows
\bea \label{cutoff relation}
\widetilde{\Lambda}^2=\frac{\rho(y)}{\bar{\rho}}\Lambda^2.
\eea
This makes sense provided $\rho(y)/\bar{\rho}$ varies only on large scales respectively small momenta $p\ll \Lambda$, which is our 
working assumption. We 
obtain as a final result for the geometric
one-loop effective action 
expressed in terms of gauge theory on $\R^4_\theta$
\bea
\Gamma_{\Psi}&=&-4 \Gamma_{\Phi}-\int d^4x \bar{\rho} \frac{\Lambda^2}{2}\bar{g}^{ac}\bar{g}^{bd}\bar F_{ab}\bar{\partial}^2 \bar F_{cd} \nn\\
&=&-4 \Gamma_{\Phi}
+\int \frac{d^4 p}{(2\pi)^4}  \widetilde{g}^{ac}\widetilde{g}^{bd}\bar F_{ab}(p) \bar F_{cd}(-p)\frac{p^2}{\Lambda_{NC}^{4}}\frac{\Lambda^2}{2}
\label{S-eff-geom-expand}
\eea
where $p^2 = p_i p_j g^{ij}$. This agrees 
precisely with the one-loop computation in the
gauge theory point of view  obtained below.
Note that the last term corresponds to $\tr\,\cE$
in \eq{induced-susy-E}.

\section{Comparison with UV/IR mixing}
\label{sec:UV-IR-mixing}

In this section, we compare the geometrical form of the one-loop
effective action obtained in the
previous section with the one-loop effective action obtained from
the gauge theory point of view. The strategy is to apply first the concept of covariant coordinates to obtain a noncommutative gauge theory coupled to fermions and compute thereafter the one-loop effective action. The result is of course the same, 
which provides not only a nontrivial check for our geometrical 
interpretation, but also sheds new light on the conditions
to which extent the semi-classical analysis of the previous section is
valid. This generalizes the results of \cite{Grosse:2008xr} 
to the fermionic case.
We find as expected that the UV/IR mixing terms obtained by
integrating out the fermions are given by the 
induced geometrical resp. 
gravitational action \eq{S-oneloop-fermions}, in a suitable IR regime.
In particular, we need an explicit, physical momentum cutoff $\Lambda$.

Using the variables and conventions of the previous section,
the action \eq{fermionic-action-geom} can be exactly 
rewritten as $U(1)$ gauge theory on $\R^4_\theta$,
which in the Euclidean case takes the form
\bea
S[\Psi] &=& (2\pi)^2\,  \Tr \Psi^\dagger \gamma_a [Y^a,\Psi]  \nn\\
&=& \int  d^4 x\, \tilde\Psi^\dagger
i\tilde \gamma^a (\bar\partial_a\tilde\Psi +i g[A_a, \tilde\Psi])
\eea
We introduce an explict coupling constant
$g$, and define a rescaled fermionic field 
\be
\tilde \Psi = |\bar g_{ab}|^{\frac 1{16}}\,\Psi
\label{psi-tilde}
\ee
in order to
obtain the properly normalized effective metric $\tilde g^{ab}$;
we will omit 
the tilde on $\Psi$ henceforth. Recall also that
only $U(1)$ gauge fields are considered here, 
because only those correspond to the nontrivial geometry
considered in the previous section. 

We need the $O(A^2)$ contribution to the one-loop effective action
obtained by integrating out the fermionic field $\Psi$.
While this  computation has been discussed several times in the 
literature \cite{Minwalla:1999px,Hayakawa:1999zf,Matusis:2000jf,VanRaamsdonk:2001jd,Khoze:2000sy}, 
the known results are 
not accurate enough for our purpose, i.e. 
in the regime $p^2, \Lambda^2<\Lambda_{NC}^2$ 
where the semiclassical geometry is 
expected to make sense. 
We need to analyze carefully the  IR regime of
the well-known effective cutoff $\Lambda_{eff}(p)$ \eq{lambda-eff} 
for non-planar graphs as $p \to 0$, keeping $\Lambda$ fixed. 
In this regime the non-planar diagrams 
almost coincide with the planar diagrams, and the leading IR corrections
due to the nonplanar diagrams correspond 
to the induced geometrical terms in \eq{S-oneloop-fermions}.
This has not been considered in 
previous attempts to explain UV/IR mixing, e.g. in terms of
exchange of closed string modes \cite{Armoni:2001uw,Sarkar:2005jw}.

\begin{figure}[t]
\begin{center}
\includegraphics[scale=0.7]{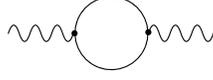}
\end{center}
\caption{Fermionic one-loop diagram.}
\label{fig:1}
\end{figure}

To proceed we use the fermionic Feynman rules and consider the Feynman 
diagram in Figure \ref{fig:1} 
corresponding to
\bea
\Gamma_{\Psi} &=& - \frac 12 \Tr \log \Delta_0 
-\frac{g^2}2 \left<\int d^4 x\, \bar\rho\, \bar \Psi \tilde\gamma^a [A_a, \Psi] 
\int d^4 y\, \bar\rho\, \bar \Psi \tilde\gamma^b [A_b, \Psi]\right>  \nn\\
&=& -\frac 12 \Tr \log \Delta_0 +\Gamma_{\Psi}(A) .
\eea
The minus sign in front is due to the fermionic loop.
This integral looks explicitly as follows
\bea
\Gamma_{\Psi} &=& - 4 g^2\int \frac{d^4 p}{(2\pi)^4}\, 
A_{a'}(p) A_{b'}(-p)\,\tilde g^{a'a} \tilde g^{b'b}\, \int \frac{d^4 k}{(2\pi)^4}\,
\frac{2k_a k_b + k_a p_b + p_a k_b - \tilde g_{ab} k(k+p)}
{(k\cdot k)((k+p)\cdot(k+p))}\, \nn\\
&& \Big(e^{-i k_i \theta^{ij} p_j}-1\Big)
\label{Gamma-fermi}
\eea
which is quite close to the bosonic case, using the notation
\bea
k\cdot k \equiv k_i k_j\, \tilde g^{ij}   \qquad
k^2 \equiv k_i\, k_j g^{ij} \,. 
\label{norm-notation}
\eea
An evaluation of the integral gives
\bea
\Gamma_{\Psi} 
&=& - 4\, \Gamma_{\Phi} 
- g^2 n_f\,\int \frac{d^4 p}{(2\pi)^4}\, 
A_{a'}(p) A_{b'}(-p)\,\tilde g^{a'a} \tilde g^{b'b}\,
 (p_a p_b - \tilde g_{ab} p\cdot p) \,\nn\\
&& \frac 1{8 \pi^2}\, \int_0^1 dz\, 
\Big(K_0(2\sqrt{\frac{z(1-z)p\cdot p}{\Lambda^2}})
 - K_0(2\sqrt{\frac{z(1-z)p\cdot p}{\Lambda_{eff}^2}}) \Big) \,,
\label{Gamma-fermi-3}
\eea
for Dirac fermions, where
\bea
\Lambda_{eff}^2 &=& { 1 \over 1/\Lambda^2 + \frac 14 \frac{p^2}{\Lambda_{NC}^4}} 
= \Lambda_{eff}^2(p) 
\label{lambda-eff}
\eea
is the ``effective'' cutoff for non-planar graphs, and $\Lambda_{NC}$ is
defined in \eq{effective-metric-bar}.
To proceed we consider the IR regime  
\be
\frac{p^2 \Lambda^2}{\Lambda_{NC}^4} < 1 \, .
\label{IR-regime}
\ee
Then both $\Lambda$ and $\Lambda_{eff}$ are large, and we can 
use an asymptotic expansions for the Bessel function
\be
K_0\Big(2 \sqrt{\frac{m^2}{\Lambda^2}}\Big)  
\,=\,  -  \(\gamma + \log(\sqrt{\frac{m^2}{\Lambda^2}})\) 
\, + O\Big(\frac{m^2}{\Lambda^2}\log(\frac{\Lambda}{m})\Big) 
\label{bessel-expand}.
\ee
Moreover, in the valid regime $p\Lambda < \Lambda_{NC}^2$ one is allowed to expand the effective cutoff
\bea\label{lambda-eff-expand-1}
\Lambda_{eff}^2 = \Lambda^2 - p^2 \frac{\Lambda^4 }{4\Lambda_{NC}^4} +... \, ,
 \qquad
\Lambda_{eff}^4 = \Lambda^4 - p^2 \frac{\Lambda^6 }{2\Lambda_{NC}^4} +... \, .
\eea
We obtain our final result
\bea
\Gamma_{\Psi} + 4\, \Gamma_{\Phi} 
&\sim&  \, \frac 14  \,\frac{g^2}{16\pi^2} \int \frac{d^4 p}{(2\pi)^4}\, 
 \tilde g^{a'a} \tilde g^{b'b}\, \bar F_{ab}(p) \bar F_{a'b'}(-p)\,
 \frac{p^2\Lambda^2}{\Lambda_{NC}^4} \, ,  \nn\\
&=&  \, \frac 14  \,\frac{g^2}{16\pi^2} \int \frac{d^4 p}{(2\pi)^4}\, 
 \bar\rho^2\Lambda^2  p^2 \,\bar g^{a'a}\bar g^{b'b}\, 
 \bar F_{ab}(p) \bar F_{a'b'}(-p)\, ,
\label{ind-action-gauge-boson-fermion}
\eea
where $p^2 = p_a p_b g^{ab}$. There are
obvious modifications due to the appropriate expansion of $\Lambda_{eff}^2$
if one approaches the border of the IR regime \eq{IR-regime}.

To compare this with the geometrical results, we must take into
account the  different regularizations used in the heat-kernel 
expansion \eq{heatkernel-expand}
and in the above one-loop computation. 
It was shown in \cite{Grosse:2008xr} 
that these regularizations agree if we  replace $\Lambda^2$ with $2 \Lambda^2$
in the one-loop computation above\footnote{While this was strictly
  speaking established only for the bosonic case, the argument should 
extend to the fermionic case without difficulties.}. 
We then find  complete agreement with the result 
\eq{S-eff-geom-expand} obtained using
the geometrical point of view. 
Notice in particular that the induced gravitational action 
is nontrivial even in
the case of e.g. $N=1$ supersymmetry. This is now understood 
in terms of induced gravity, and full cancellation is obtained only
in the case of $N=4$ supersymmetry. This will be discussed below.

\paragraph{Cancellations and supersymmetry}
\label{susy-cancellations}

It is very interesting to compare the fermionic and the 
bosonic contribution to the gravitational action.
As is well-known \cite{Matusis:2000jf,Khoze:2000sy}, we note that the
fermionic contribution to the 
one-loop effective action in NC gauge theory
does not quite cancel the scalar contribution, due to 
\eq{ind-action-gauge-boson-fermion}. 
This means that even in supersymmetric
cases some UV/IR mixing may remain. 
From the geometrical point of view, 
this terms corresponds to a gravitational action
$\tr\, \cE\, \tilde\Lambda^2= - 2 R[\tilde G]\, \tilde\Lambda^2 +
...$, so that the 
cutoff $\tilde\Lambda^2$  should be interpreted as  
effective gravitational constant $\frac 1G$. 
This is completely analogous to the commutative case, where the
gravitational term \eq{cE-commutative} is induced.
The remaining UV/IR mixing
term cancels only in the case of $N=4$ supersymmetry.
We can therefore identify $\tilde\Lambda$ as the scale of $N=4$ SUSY breaking
(assuming such a model), above which the model is finite. 
These observations strongly suggest that 
for the model to be well-defined at the quantum level,
$N=4$ SUSY
is required above the gravity scale i.e. the Planck scale.
This is  realized by the 
IKKT model \cite{Ishibashi:1996xs} on a NC background. 

\section{Discussion and outlook}

In this paper, fermions are studied in the framework of 
emergent noncommutative gravity, as realized through matrix models
of Yang-Mills type. The matrix model strongly suggests a
particular fermionic term in the action, 
corresponding to a specific coupling to
a  background geometry with nontrivial metric $\tilde G_{\mu\nu}$. 
This coupling is similar to the standard coupling of fermions to a 
gravitational background, except that the spin connection vanishes
in the preferred coordinates associated with the matrix model. 

The main result of this paper is that in spite of this unusual
feature, the resulting fermionic action is very reasonable,
and properly describes fermions coupled to emergent gravity. 
In the point particle limit, 
fermions propagate along the appropriate trajectories, albeit with
a different rotation of the spin. At the quantum level, we find an
induced gravitational action which includes the expected
Einstein-Hilbert term with a modified coefficient, 
as well as an additional
term for a scalar density reminiscent of a dilaton. 
There are further terms which vanish for on-shell geometries.
We conclude that the framework of emergent gravity does extend to 
fermions in a reasonable manner, and might well provide
- in a suitable extension - a physically viable theory of gravity.

In a second part of the paper, we compare this induced gravitational 
action 
with the well-known UV/IR mixing in NC gauge theory due to fermions.
Generalizing the results in \cite{Grosse:2008xr} 
for scalar fields, we find 
as expected that the UV/IR mixing can be explained precisely by the
gravitational point of view. 
This also provides a nice understanding for the  fact that 
some UV/IR mixing remains in supersymmetric cases, and only 
disappears for $N=4$ supersymmetry. The reason is that a 
gravitational action is induced even in supersymmetric cases, 
except in $N=4$ SUSY. This in turn leads to the 
conjecture that the gravitational 
constant should be related to the scale of $N=4$ SUSY breaking,
which is quite reasonable. All of these findings suggest that the 
IKKT model
on a noncommutative background 
\cite{Ishibashi:1996xs,Aoki:1999vr,Ishibashi:2000hh,Kitazawa:2005ih} 
should be the most promising candidate for a realistic version of 
emergent gravity.
These issues will be discussed in more detail elsewhere.

\paragraph{Acknowledgments}

The work of D.K. was supported by the FWF project P20017, and
the work of H.S. was supported in part by the FWF project P18657 and 
in part by the FWF project P20017.

\end{document}